\documentclass[aps,prd,showpacs]{revtex4}
\usepackage[utf8x]{inputenc}
\usepackage{fancyhdr}
\usepackage{graphics}
\usepackage{epsfig}
\usepackage{graphicx}
\usepackage{epsf}
\usepackage{enumitem}
\usepackage{amsmath, amsthm, amssymb}
\usepackage{slashed}
\usepackage{subfig}
\usepackage{nicefrac}
\usepackage{url}
\usepackage{color}
\usepackage{ulem}

\newcommand{\g}{\gamma}

\begin{document}
\title{Light Axigluon Contributions to $b\bar{b}$ and $c\bar{c}$ Asymmetry and Constraints on Flavor Changing Axigluon Currents }
\author{Seyda Ipek}
\affiliation{Department of Physics, University of Washington, Seattle, Wa 98195, USA}
\date{\today}

\begin{abstract}
The light axigluon model  is a viable candidate to explain the Tevatron $t\bar{t}$ forward-backward asymmetry. In this paper we present the forward-backward asymmetries for $b\bar{b}$ and $c\bar{c}$ systems predicted by a broad light axigluon with mass 100-400 GeV. Furthermore, we modify this flavor universal axigluon model to include flavor changing couplings of axigluons with the SM quarks. We constrain these couplings from the available neutral meson mixing data, and investigate their effects on the rare decay $B^0_s\to\mu^+\mu^-$, CP violating $D\to h^+h^-$ and isospin violating $B\to K^{(*)}\mu^+\mu^-$ decays. We show that a light axigluon can contribute to the observed CP violation in $D\to h^+h^-$.
\end{abstract}

\pacs{12.90.+b, 14.65.Ha, 14.40.-n, 11.30.Hv}
\maketitle

\section{Introduction}
In the Standard Model (SM), the process $q\bar{q}\to t\bar{t}$ is symmetric under the exchange of $t$ and $\bar{t}$ at leading order (LO). When next-to-leading order (NLO) processes are included, there is a small forward-backward asymmetry (FBA), of $A_{FB}^{SM}=0.06\pm0.01$ \cite{SMasym1,SMasym2,SMasym3,SMasym4}. This non-zero and positive asymmetry means that (anti-)top quarks are emitted preferably in the incoming (anti-)quark direction. In 2011, CDF \cite{CDF1,CDF2} and D\O\ collaborations \cite{d0} measured a high FBA in $t\bar{t}$ production from proton-antiproton collisions. The  D\O\ asymmetry is $0.196\pm0.065$ and the CDF asymmetry is $0.164\pm 0.047$. Furthermore, CDF reported a mass dependent asymmetry \cite{CDF2}:
\begin{align}
 A_{FB}^{t\bar{t}}(m_{t\bar{t}}>450\text{GeV})&=0.296\pm0.067\\
 A_{FB}^{t\bar{t}}(m_{t\bar{t}}<450\text{GeV})&=0.078\pm0.054
\end{align}

On the other hand, the charge asymmetry measured at ATLAS ($A_C=-0.019\pm0.028\pm0.024$ \cite{atlas}) and CMS ($A_C=0.004\pm0.010\pm0.011$ \cite{cms}) agrees well with the SM predictions.

There are various new physics (NP) models to explain the FB asymmetry, many of which are in tension with the LHC charge asymmetry, like sign top production, and the $t\bar{t}$ cross section. In this paper we will consider and modify  one of the light axigluon models suggested by Tavares and Schmaltz \cite{tavares}, which is still a viable candidate \cite{drob,tavares2}. Axigluons have a long history \cite{a9,a1} and there has been a significant amount of work to explain the $t\bar{t}$ FBA via massive color octets \cite{tt1,a2,a3,a4,a6,a7,a8,a10}. 

For details of the model, see \cite{tt1,tavares,tavares2}. To summarize, the model has an extra $SU(3)$ symmetry group, hence the gauge symmetry is $SU(3)_1\times SU(3)_2\times SU(2)_W \times U(1)_Y$. Introduced with this extra symmetry group is an extra set of up- and down- type quarks, and a scalar field $\Phi$, which acquires a vacuum expectation value  (\emph{vev}) to break $SU(3)_1\times SU(3)_2$ into the diagonal $SU(3)_c$ of the SM. Through this symmetry breaking, one combination of the two $SU(3)$ gauge fields acquires a mass. This massive color octet is called the axigluon, and its massless counterpart is the SM gluon. Similarly, there are combinations of fermions that become exotic heavy quarks and the SM light quarks, allowing the axigluon coupling to the light quarks to be a free parameter. Also, there are no gauge anomalies due to cancellations from the additional quarks. Gluons couple to both the SM and heavy quarks with the same strength as expected. The lepton sector is exactly the same as 
the SM, and will not be mentioned throughout this paper. The axigluon in this model can have mass below 450 GeV. However, to be viable, it needs to be broad. In \cite{tavares,tavares2}, the authors introduce new heavy quarks and color adjoint scalars that the axigluon can decay into. These exotic quarks and scalars then decay into multi-jets which is not ruled out by LHC searches yet. Note also that axigluons with mass $m>2m_t$ and enhanced couplings to top quarks can be seen via LHC four-top searches \cite{fourtops}. However this axigluon is fairly light, and it does not have enhanced top couplings.

In \cite{tavares}, the authors consider only flavor universal couplings of axigluons to the SM quarks. This relies on the strong assumption of an underlying global symmetry. This global symmetry is only approximate. Mixing of heavy and light quarks could induce flavor changing neutral currents (FCNCs). Furthermore, since the mixing occurs between quarks that have the same $SU(2)\times U(1)$ charge, it does not give rise to flavor changing $Z$ couplings. Therefore we will not assume the existence of an exact global symmetry of the axigluon couplings, which allows flavor changing couplings of the axigluons. The new scalars in this model do not induce FCNCs, so the axigluon couplings are the most significant possible source of new FCNC. Other models that have flavor changing color-octet couplings have been proposed in the literature \cite{faisel,Haisch}. A general Lagrangian with flavor violating axigluon interactions contains the following terms: 
\begin{align}
 \mathcal{L}\supset\hspace{0.05in} &\bar{u}_i\gamma_\mu A^\mu (g_{uL}^i\delta_{ij}+(\epsilon^u_L)_{ij})P_Lu_j+\bar{d}_i\gamma_\mu A^\mu (g_{dL}^i\delta_{ij}+(\epsilon^d_L)_{ij})P_Ld_j \notag \\
& +\bar{u}_i\gamma_\mu A^\mu (g_{uR}^i\delta_{ij}+(\epsilon^u_R)_{ij})P_Ru_j+\bar{d}_i\gamma_\mu A^\mu (g_{dR}^i\delta_{ij}+(\epsilon^d_R)_{ij})P_Rd_j
\end{align}
Here $A^\mu$ is the axigluon, $u_i$ and $d_i$ are SM up- and down-type quarks respectively ($i=1,2,3$ is the generation index), and $g^i$ are flavor independent couplings. Color and spinor indices are suppressed for simplicity. The complex matrices
\begin{align} \label{fcc}
 \epsilon^d_{L,R}=\left(\begin{array}{ccc}  
         0 & g_{ds}^{L,R} & g_{db}^{L,R}\\
	 {g_{ds}^{L,R}}^* & 0 & g_{bs}^{L,R} \\
	 {g_{db}^{L,R}}^* & {g_{bs}^{L,R}}^* & 0
         \end{array} \right), \hspace{0.5in}
 \epsilon^u_{L,R}=\left(\begin{array}{ccc}
         0 & g_{uc}^{L,R} & g_{ut}^{L,R}\\
	 {g_{uc}^{L,R}}^* & 0 & g_{ct}^{L,R} \\
	 {g_{ut}^{L,R}}^* & {g_{ct}^{L,R}}^* & 0
         \end{array} \right)
\end{align}
contain off diagonal axigluon couplings of up- and down-quarks respectively. This mixing follows from flavor symmetry breaking of heavy and light quarks. These FCNCs, which can occur at tree level, can have interesting effects on FCNC observables. 

The rest of the paper is organized as follows: In Section 1, we investigate the contribution of light axigluons to $b\bar{b}$ and $c\bar{c}$ FBAs. In Section 2, we constrain the flavor changing axigluon couplings from neutral meson mixing data. In Section 3, we investigate the contribution of the constrained flavor changing axigluon model to the following decays: $B^0_s\to\mu^+\mu^-$, $D^0\to h^+h^-$, and $B\to K^{(*)}\mu^+\mu^-$.

\section{Forward-Backward Asymmetries}

\begin{figure}[b]
\begin{center}
 \includegraphics[scale=0.8]{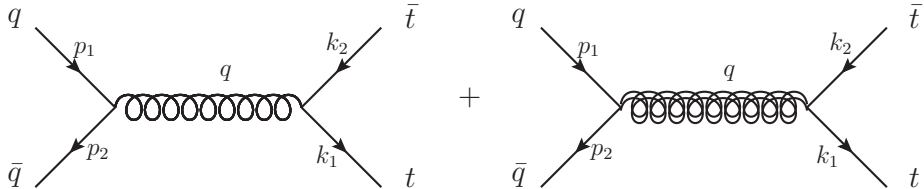}
 \caption{Diagrams contributing to $q\bar{q}\to t\bar{t}$ amplitude. Left diagram is from the SM gluon exchange, right diagram is the axigluon exchange.}\label{tta}
\end{center}
\end{figure}

In the Tevatron experiments CDF and D\O\ , $t\bar{t}$ production happens mostly via $q\bar{q}\to t\bar{t}$ (Fig.\ref{tta}).  The square amplitude of the process $q\bar{q}\to t\bar{t}$ including the axigluon contribution is calculated as follows \cite{tavares,tt1}:
\begin{align} \label{ttax}
 |\mathcal{M}|^2&=Ng_s^4(1+c^2+4m^2)\notag \\
&\hspace{10pt}-Ng_s^4\frac{2\hat{s}(-\hat{s}+M_A^2)}{(-\hat{s}+M_A^2)^2+\Gamma_A^2M_A^2}\biggl[g_V^tg_V^q(1+c^2+4m^2)+2g_A^tg_A^q\,c\biggr] \notag\\
&\hspace{10pt} +Ng_s^4\frac{\hat{s}^2(-\hat{s}+M_A^2)^2}{((-\hat{s}+M_A^2)^2+\Gamma_A^2M_A^2)^2}\biggl[\bigl[g_V^{q\,2}+g_A^{q\,2}\bigr]\bigl[g_V^{t\,2}(1+c^2+4m^2)+g_A^{t\,2}(1+c^2-4m^2)\bigr]+8g_V^tg_A^tg_v^qg_A^q\,c\biggr]
\end{align}
where $N=\frac{4}{9}$ is the color sum, $\hat{s}=-(p_1+p_2)^2$ is the partonic total momentum, $\beta=\sqrt{1-{4m_t^2\over \hat{s}}}=\sqrt{1-4m^2}$ is the velocity of the top quark, $c\equiv\beta\cos\theta$ where $\theta$ is the angle between the incoming $q$- and the outgoing $t$-quark, $M_A$ is the axigluon mass, and $\Gamma_A$ is the width. Vector and axial couplings of the axigluon are defined as:
\[ g_V^q=\frac{g_R^q+g_L^q}{2}, \hspace{10pt} g_A^q=\frac{g_R^q-g_L^q}{2} \]
In Eq.\ref{ttax}, the first term comes from the SM gluon exchange, the second term is the interference  between the gluon and the axigluon channels, and the third term is the axigluon s-channel (See Fig.\ref{tta}). The FBA comes from the terms that are proportional to the odd powers of $\cos\theta$, since $\cos\theta$ is odd under $\theta\to \pi-\theta$. In order to accommodate the measured $t\bar{t}$ FBA, the axigluon should give a large FBA without affecting the $t\bar{t}$ cross-section, which is close to its SM value. A light axigluon ($M_A = 100-400$ GeV) with a large width $\Gamma_A\simeq0.1M_A$, is shown to agree with both the $t\bar{t}$ FBA and the $t\bar{t}$ cross section \cite{tavares,tavares2}. Assuming approximate parity symmetry, $g_V^i$ needs to be small. We will, as in \cite{tavares}, take $g_V^i=0$. 

\begin{figure}[h]
\centering
\subfloat[Integrated $A_{FB}$ vs $t\bar{t}$ invariant mass $m_{t\bar{t}}$.]{
\includegraphics[scale=0.8]{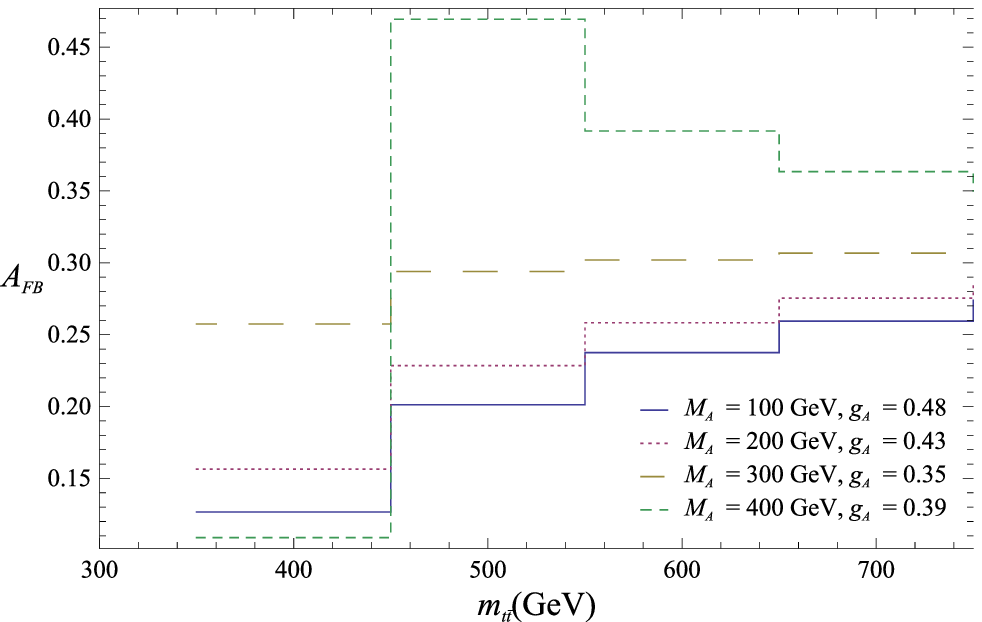} 
}
\subfloat[CDF graph of $A_{FB}$ vs $t\bar{t}$ invariant mass $m_{t\bar{t}}$ \cite{CDF2}.]{
\includegraphics[scale=0.8]{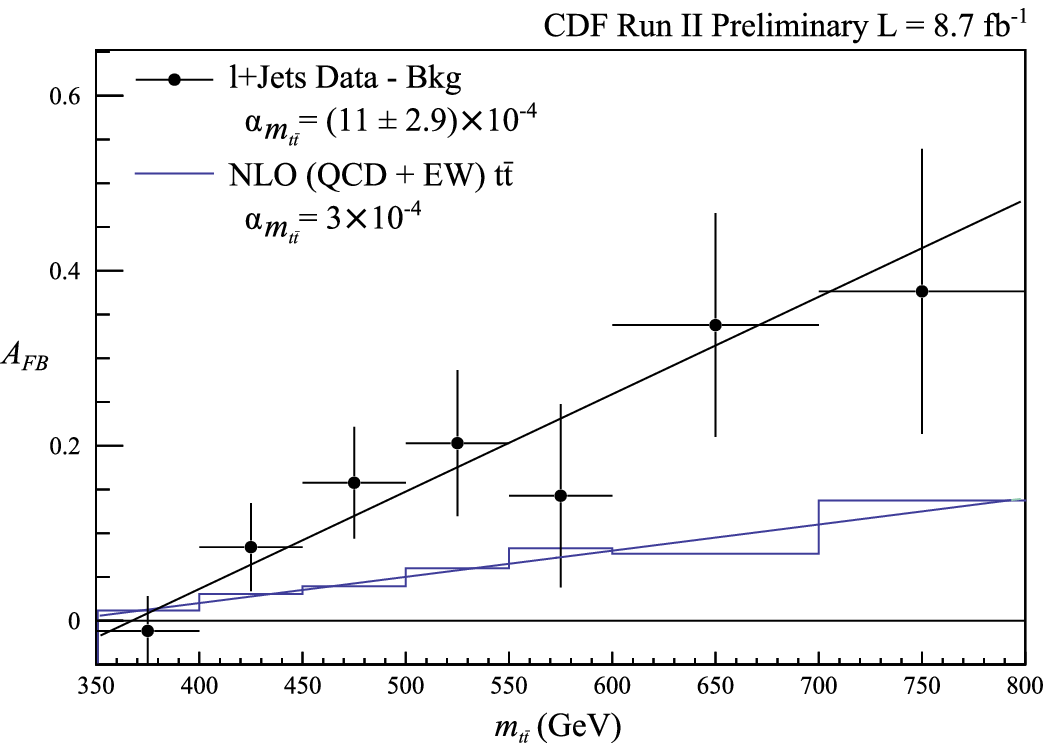} 
}
\caption{Comparison of the light axigluon prediction and the recent CDF measurement of $t\bar{t}$ $A_{FB}$. Due to the large error in the data, $M_A{=}400$ GeV is not precluded. The values for the coupling constant $g_A$ are taken from \cite{tavares2}.} \label{tasym}
\end{figure}

We will define the FBA, $A_{FB}$, through the forward and backward scattering cross-sections:
\begin{align*}
 \sigma_+&=\sigma(0<\theta<\pi/2) \\
 \sigma_-&=\sigma(\pi/2<\theta<\pi)
\end{align*}
where $\sigma=\frac{\beta}{32\pi \hat{s}}\int d\cos\theta\,|\mathcal{M}|^2$ is the total cross-section. Then the FBA is
\begin{align}
 A_{fb}&=\frac{\sigma_+-\sigma_-}{\sigma_+ + \sigma_-}\\
&=-\frac{\beta g_A^2\frac{-\hat{s}+M_A^2}{(-\hat{s}+M_A^2)^2+\Gamma_A^2M_A^2}}{\frac{1}{2\hat{s}}\bigl(1+{\beta^2\over3}+4m^2\bigr)+\frac{\hat{s}(-\hat{s}+M_A^2)^2}{2\bigl[(-\hat{s}+M_A^2)^2+\Gamma_A^2M_A^2\bigr]^2}g_A^4\bigl(1+{\beta^2\over3}-4m^2\bigr)}\label{fba}
\end{align}
where $g_A^q=g_A^t=g_A$. This FBA is plotted in Fig.\ref{tasym} as a function of the $t\bar{t}$ invariant mass $m_{t\bar{t}}$. The values for coupling constants for each axigluon mass (100, 200, 300, 400 GeV) are taken from  \cite{tavares2}, where the authors performed fits to both the Tevatron FBA and the LHC charge asymmetry, and chose the coupling constants that would best fit both of them. 

\begin{figure}[b]
\begin{center}
 \includegraphics[scale=0.8]{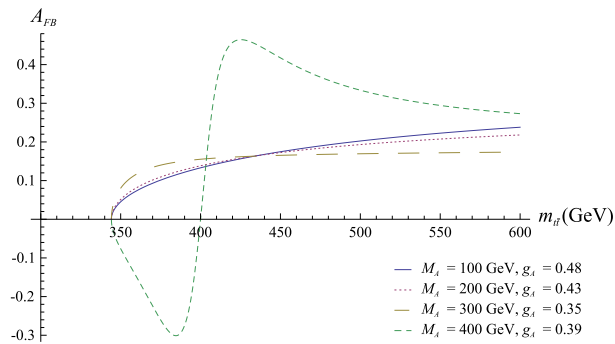}
 \caption{$t\bar{t}$ forward backward asymmetry, $A_{FB}$ vs $t\bar{t}$ invariant mass $m_{t\bar{t}}$. $A_{FB}$ only changes sign when $M_A>2m_t$. The values for the coupling constant $g_A$ are taken from \cite{tavares2}.}
\end{center}
\end{figure}

\begin{figure}[h]
\centering
\subfloat[$b\bar{b}$ forward-backward asymmetry vs $b\bar{b}$ invariant mass $m_{b\bar{b}}$]{
\includegraphics[scale=0.8]{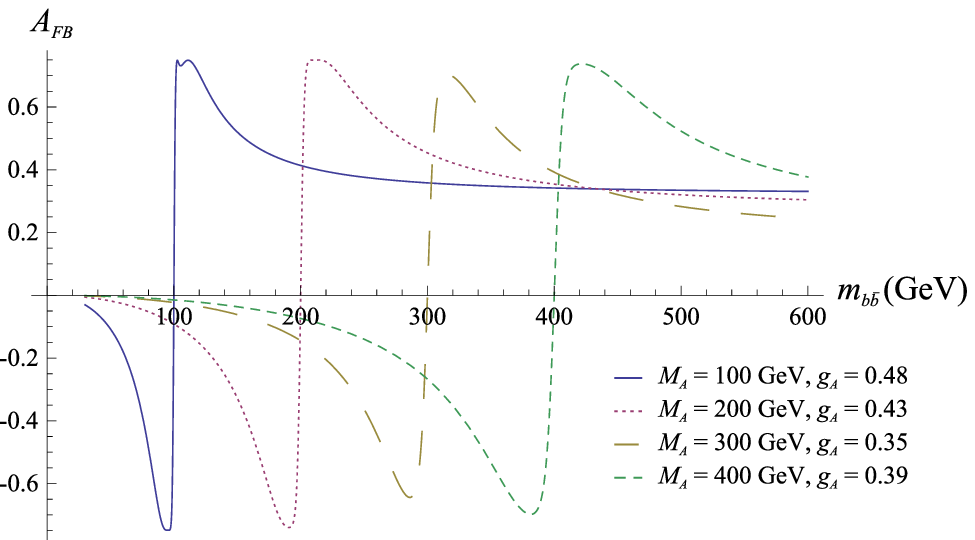}
\label{basym}
}
\subfloat[$c\bar{c}$ forward-backward asymmetry vs $c\bar{c}$ invariant mass $m_{c\bar{c}}$]{
\includegraphics[scale=0.8]{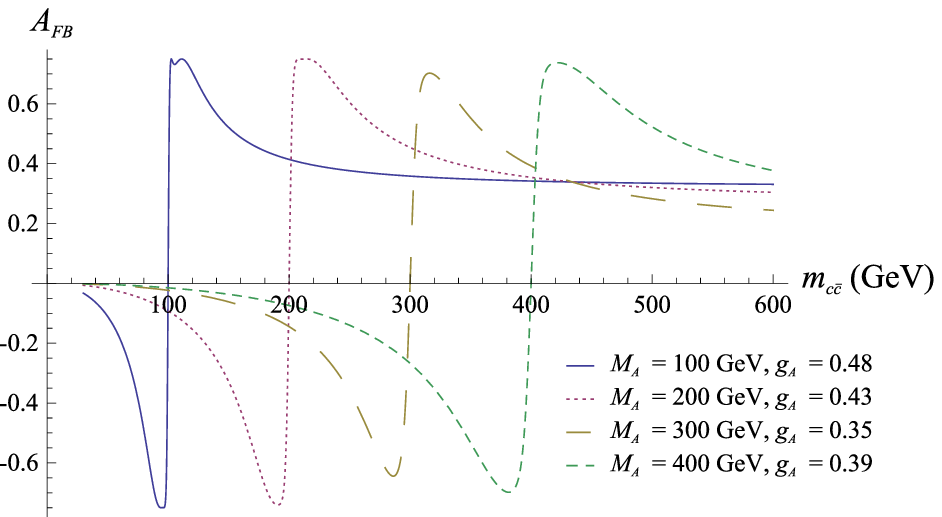}
 \label{casym}
}
\caption{Forward-backward asymmetries for bottom and charm quarks vs $q\bar{q}$ invariant mass $m_{q\bar{q}}$, for $q=b,c$. All asymmetries cross zero when $m_{q\bar{q}}=M_A$ which is much higher than $m_{q\bar{q}}=2m_q$. The values for the coupling constant $g_A$ are taken from \cite{tavares2}. }\label{nasym}
\end{figure}

An obvious prediction of the light axigluon model is FBAs for bottom and charm quarks. Although these may be challenging to measure, their observation would be an important clue. The graphs for these asymmetries can be seen in Fig.\ref{basym} and Fig.\ref{casym}. Since both $m_b=4.2$ GeV and $m_c=1.27$ GeV are very small compared to the predicted axigluon mass (100s GeVs), the FBA structure is almost the same for both quarks. As can be seen from Fig.\ref{nasym}, the predicted FBAs are quite large even for low energies, $m_{b\bar{b}}\simeq$ 200 GeV, so the search need\sout{s} not \sout{to} go to high invariant masses. Furthermore, the crossing at $m_{q\bar{q}}=M_A$ ($q=b,c$) is expected to be clearer compared to $t\bar{t}$ production, since $M_A>>2m_q$. Consequently, measuring the $b\bar{b}$ FBA  would be a good way to find out the axigluon mass. Measuring the $b\bar{b}$ FBA is also suggested in \cite{basym1}. 

\begin{figure} [h]
\centering
\subfloat[$b\bar{b}$ forward-backward asymmetry vs $b\bar{b}$ invariant mass $m_{b\bar{b}}$]{
\includegraphics[scale=0.8]{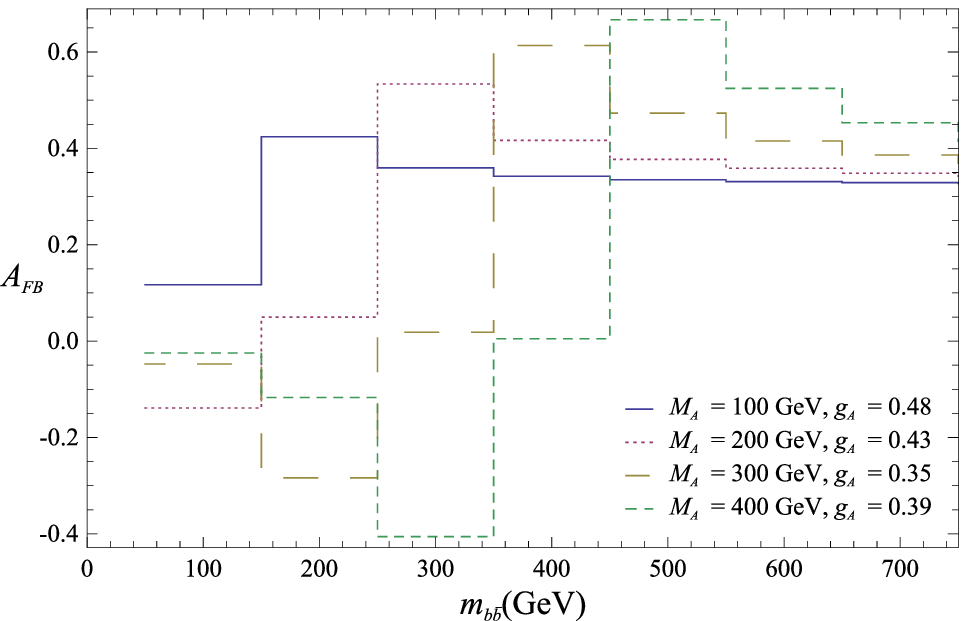}
\label{basym2}
}
\subfloat[$c\bar{c}$ forward-backward asymmetry vs $c\bar{c}$ invariant mass $m_{c\bar{c}}$]{
\includegraphics[scale=0.8]{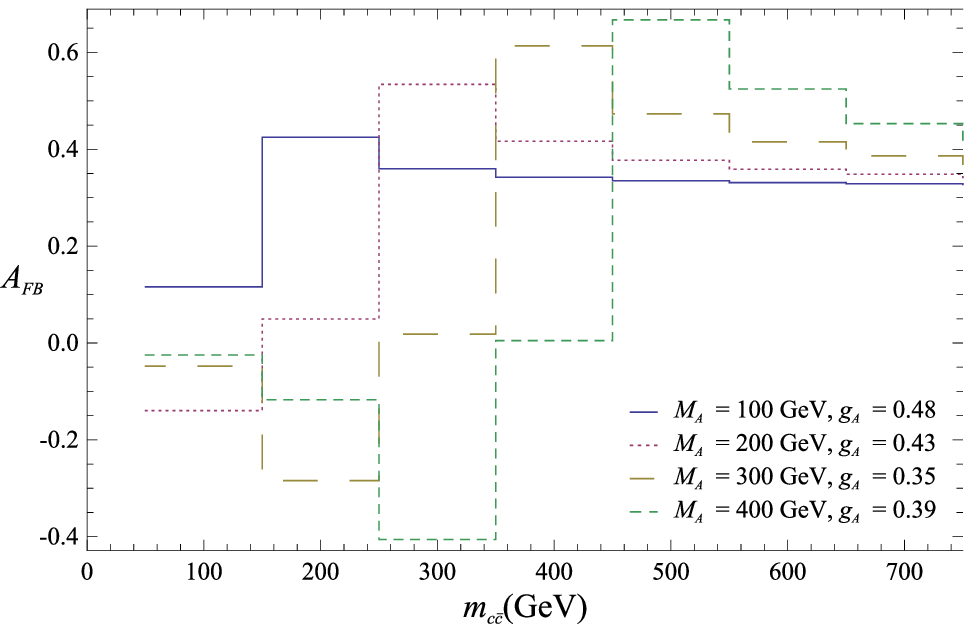}
 \label{casym2}
}
\caption{Integrated asymmetries from Fig.\ref{nasym}} \label{intasym}
\end{figure}

\section{Constraints on Flavor Changing Axigluon Currents From Neutral Meson Mixings}

\begin{figure}
\begin{center}
 \includegraphics[scale=0.8]{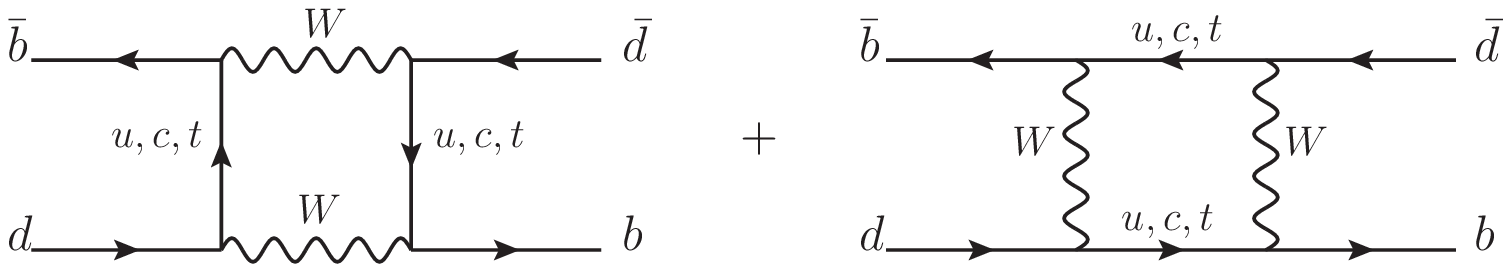}
 \caption{Electroweak box diagrams that contribute to $B^0-\bar{B^0}$ mixing}\label{ewbox}
\end{center}
\end{figure}

In this section we will use the data available from neutral meson mixings ($B^0_q-\bar{B}^0_q, K^0-\bar{K}^0, D^0-\bar{D}^0$ mixing) to constrain the FC axigluon coupling matrices, $ \epsilon^u_{L,R}$ and $ \epsilon^d_{L,R}$, in Eq.\ref{fcc}. Neutral mesons ``mix" because the flavor eigenstates ($M^0$, $\bar{M^0}$) of the SM Hamiltonian are not the actual mass eigenstates ($M_{L,H}$, $L$ and $H$ for light and heavy respectively). The mass difference between $M_L$ and $M_H$ will be one of the constraints we will use for each mixing. Since these processes are FCNC processes they happen via loop diagrams in the SM, like electroweak (EW) box diagrams in Fig.\ref{ewbox}. These diagrams are suppressed by Cabibbo-Kobayashi-Maskawa (CKM) matrix elements due to flavor changing EW interactions. However, with flavor changing axigluon couplings  we have neutral meson mixing at tree level. Therefore the FC couplings are constrained by the mixing data for $B^0_q-\bar{B}^0_q$ and $K^0-\bar{K^0}$ and $D^0-\bar{D^0}$ mixing.

For the following calculations we take $M_A=400$ GeV, which gives us the least stringent constraints. We explain the methods we use to constrain each coupling constant in the respective sections. In each section we try to give the most general forms that can be used to constrain these couplings, but at the end we assume axial couplings for simplicity. We prefer to give constraints on real and imaginary parts of $g_{ij}^2$ ($i,j=$ quark flavors) rather than $g_{ij}$ itself, since the meson mixing amplitudes that we use for the constraints involve the square of the coupling constants. One can find the constraints on the real and imaginary parts of the coupling constants themselves, assuming neither of them are zero. A summary of the results can be seen in Table. \ref{table} .

\begin{table}[h] 
 \begin{tabular}{|c|c|c|c|}
  \hline
  Coupling & Constraint &  Im($g_{ij}^2$) & $|g_{ij}|$ \\ \hline
  $g_{bd}$ & $B^0-\bar{B}^0$ & $<2.10\times10^{-7}$ & $<4.58\times10^{-4}$ \\ \hline
  $g_{bs}$ & $B^0_s-\bar{B}^0_s$ & $<2.55\times10^{-6}$ & $<1.83\times10^{-3}$ \\ \hline
  $g_{ds}$ & $K^0-\bar{K^0}$ & $<6.13\times10^{-13}$ & $<3.11\times10^{-5}$ \\ \hline \hline
  $g_{uc}$ & $D^0-\bar{D}^0$ & $<4.89\times10^{-9}$ & $<1.47\times10^{-4}$\\ \hline
 \end{tabular}
\caption{Origin of constraints and upper bounds on the imaginary part and the magnitude of FC axigluon couplings} \label{table}
\end{table}

\subsection{\texorpdfstring{$B^0_q-\bar{B}^0_q$}{B-meson} Mixing}

\begin{figure}[b]
\begin{center}
 \includegraphics[scale=0.8]{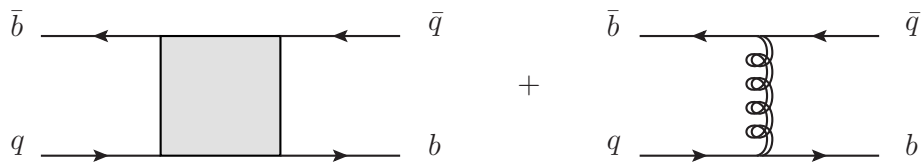}
 \caption{The SM and the axigluon contribution to $B_q-\bar{B}_q$ mixing}\label{bbmixing}
\end{center}
\end{figure}
The SM and the axigluon contribution to $B_q^0-\bar{B}_q^0$ mixing amplitude, where $q=s,d$ for $B_s^0$ and $B^0$ respectively, can be seen in Fig.\ref{bbmixing}. The gray box in the first diagram is the sum of all EW box diagrams that have $u,c,t$-quarks in the loop. However, for $B^0_q-\bar{B}^0_q$ mixing, the $t$-loop is the most important one. The SM contribution, including NLO corrections, to this amplitude is given in \cite{burasnlo} as:
\begin{align} \label{smbbar1}
 \mathcal{M}_{SM}=-i\frac{G_F^2M_W^2}{16\pi^2}(V_{tb}^*V_{tq})^2(0.551)S_0(x_t)(\alpha_s(\mu_b))^{-6/23}\left[1+\frac{\alpha_s(\mu_b)}{4\pi}J_5\right](\bar{b}q)_{V-A}(\bar{b}q)_{V-A}
\end{align}
where $G_F$ is the Fermi constant, $M_W=80.4$ GeV is the W-boson mass, $V_{ij}$ are the CKM matrix elements that mix $i-$ and $j-$type quarks, $\alpha_s(\mu_b)$ is the strong structure constant evaluated at $\mu_b\simeq \mathcal{O}(m_b)$, and $J_f$ ($\simeq 1.627$ for $f=5$, $f$ being the number of active flavors at the mixing scale) is a constant that comes from the running of the coupling coefficients. Also $(\bar{b}q)_{V-A}\equiv\bar{b}\gamma^\mu(1-\gamma_5)q$, and
\begin{align}
S_0(x_t)=\frac{4x_t-11x_t^2+x_t^3}{4(1-x_t)^2}-\frac{3x_t^3}{2(1-x_t)^3}\ln x_t
\end{align}
with $x_t={m_t^2\over M_W^2}$. This function $S_0(x)$ is one of many similar functions called Inami-Lin functions. They are loop functions that arise in box and penguin diagrams in the SM and were calculated by Inami and Lin in \cite{inami}. For $m_t=172.1$ GeV and $M_W=80.4$ GeV, $S_0(x_t)\simeq 2.51$. Thus the SM mixing amplitude is:
\begin{align}\label{smbbar2}
 \mathcal{M}_{SM}\simeq -i\frac{G_F^2M_W^2}{16\pi^2}(2.14)(V_{tb}^*V_{tq})^2(\bar{b}q)_{V-A}(\bar{b}q)_{V-A}
\end{align}

The second diagram in Fig.\ref{bbmixing} is the axigluon contribution to the mixing amplitude. This part can be written as:
\begin{align}\label{baxmix}
 \mathcal{M}_{ax}={-i\over M_A^2}\biggl\{-{C_0\over2}&\left((g_{bq}^L)^2+(g_{bq}^R)^2\right)\left(\frac{N_c+1}{N_c}\right)(\bar{b}\gamma^\mu P_L q)(\bar{b}\gamma^\mu P_L q)\notag \\
&+g_{bq}^Lg_{bq}^R\left[-{C_1\over N_c}(\bar{b}\gamma^\mu P_L q)(\bar{b}\gamma^\mu P_R q)+C_2(\bar{b} P_L q)(\bar{b}P_R q)\right]\biggr\}
\end{align}
where $N_c$ is the number of colors, and $C_0$, $C_1$ and $C_2$ are the renormalization group (RG) evolved coefficients for the corresponding 4-quark operators. These coefficients can be calculated by following \cite{adm}.

Let us define the following quantity:
\begin{align} \label{delta}
 \Delta_q=\frac{\langle B_q^0|\mathcal{H}_{SM}+\mathcal{H}_{ax}|\bar{B}_q^0\rangle}{\langle B_q^0|\mathcal{H}_{SM}|\bar{B}_q^0\rangle}
\end{align}
in order to compare the new physics (NP) and the SM contributions to the mixing process. We need the following matrix elements \cite{faisel}:
\begin{subequations}\label{meb}
\begin{align} 
 &\langle\bar{B}_q^0|\bar{b}\gamma^\mu(1\pm\gamma_5)q\bar{b}\gamma^\mu(1\pm\gamma_5)q|B_q^0\rangle = {4\over3}m_{_{B_q^0}}f_{B_q}^2\hat{B}_q \\
 &\langle\bar{B}_q^0|\bar{b}\gamma^\mu(1+\gamma_5)q\bar{b}\gamma^\mu(1-\gamma_5)q|B_q^0\rangle = -{5\over3}m_{_{B_q^0}}f_{B_q}^2\hat{B}_{1q}^{RL} \\
 &\langle\bar{B}_q^0|\bar{b}_\alpha\gamma^\mu(1+\gamma_5)q_\beta\bar{b}_\beta\gamma^\mu(1-\gamma_5)q_\alpha|B_q^0\rangle = -{7\over3}m_{_{B_q^0}}f_{B_q}^2\hat{B}_{2q}^{RL}
\end{align}
\end{subequations}
where $m_{_{B_q^0}}$ is the $B_q^0$ meson mass, $f_{B_q}$ is the decay constant, and $\hat{B}$ the bag parameter ($\hat{B}_q\sim\hat{B}_{1q}^{RL}\sim\hat{B}_{2q}^{RL}\sim1$ for B-meson decays). Recent lattice-calculated values for these constants can be found in \cite{latt, latt2} and references therein. Using Eq.\ref{meb} in Eq.\ref{smbbar2} and \ref{baxmix} we can write the SM and axigluon matrix elements for the mixing as follows:
\begin{align}
 \langle B_q^0|\mathcal{H}_{SM}|\bar{B}_q^0\rangle=&\frac{G_F^2M_W^2}{12\pi^2}(2.14)(V_{tb}^*V_{tq})^2 m_{_{B_q^0}}f_{B_q}^2 \\
\langle B_q^0|\mathcal{H}_{ax}|\bar{B}_q^0\rangle=& {1\over M_A^2}\left[-{2C_0\over9}\left((g_{bq}^L)^2+(g_{bq}^R)^2\right)+\left({5C_1\over36}-{7C_2\over24}\right)g_{bq}^Lg_{bq}^R \right]m_{_{B_q^0}}f_{B_q}^2 \notag \\
&\simeq \frac{m_{_{B_q^0}}f_{B_q}^2 }{M_A^2}\left(-0.18((g_{bq}^L)^2+(g_{bq}^R)^2)-0.73\,g_{bq}^Lg_{bq}^R\right)
\end{align}

Thus Eq.\ref{delta} reads:
\begin{align} \label{deltaq}
\Delta_q=1+\frac{12\pi^2\left(-0.18((g_{bq}^L)^2+(g_{bq}^R)^2)-0.73\,g_{bq}^Lg_{bq}^R\right)}{G_F^2M_W^2M_A^2(2.14)(V_{tb}^*V_{tq})^2}
\end{align}

As in the flavor conserving part of the light axigluon model, we assume axial couplings: $g^L_{bq}=-g^R_{bq}=g_{bq}$:
\begin{align}\label{deltaq2}
 \Delta_q=1+\frac{12\pi^2(0.37)(g_{bq})^2}{(2.14)G_F^2M_W^2M_A^2(V_{tb}^*V_{tq})^2}
\end{align}
In \cite{bbdata} one can find Re($\Delta_q$) and Im($\Delta_q$) for $B^0_s$ and $B^0$ mixing. In the next two subsections we are going to look at both cases separately.

\subsubsection{\texorpdfstring{$B^0_s-\bar{B}^0_s$}{Bs} mixing}
In this paper we use the CKM basis given in \cite{pdg}. In this basis, the relevant CKM matrix elements, $V_{tb}$ and $V_{ts}$, are real, which makes the SM contribution real. Hence, the real and imaginary parts of $g_{bs}^2$ can be constrained separately from the real and imaginary parts of $\Delta_s$.  From \cite{bbdata}, we take the following boundaries:
\begin{align}
 0.85\leq\text{Re}(\Delta_s)\leq 1.27, \hspace{0.5in} |\text{Im}(\Delta_s)|\leq 0.32
\end{align}
To be conservative, all parameters are taken at the $3\sigma$ boundaries of the fits from \cite{bbdata}, since there is some tension with the SM at $2\sigma$. Using these values and Eq.\ref{deltaq2} with $q=s$, we get the following constraints:
\begin{subequations}
\begin{align}
& |\text{Re}(g_{bs}^2)|< 2.15\times 10^{-6}, \hspace{0.3in}  |\text{Im}(g_{bs}^2)|< 2.55\times 10^{-6} \\
&\hspace{0.6in}\Longrightarrow |g_{bs}| < 1.83\times 10^{-3}
\end{align}
\end{subequations}

\subsubsection{\texorpdfstring{$B^0-\bar{B}^0$}{Bd} mixing}
For $B^0-\bar{B}^0$ mixing, one of the relevant CKM matrix elements, $V_{td}$, is complex, therefore the real and imaginary parts of $g_{bd}^2$ can not be constrained separately. Let us work this through in more detail. We can write Eq.\ref{deltaq2} as follows:
\begin{align}
 \Delta_d=1+Cg_{bd}^2
\end{align}
where 
\[C=\frac{12\pi^2(0.37)}{(2.14)G_F^2M_W^2M_A^2(V_{tb}^*V_{td})^2}=1.86+1.72\,i\]
Notice that the real and imaginary parts of $C$ are very similar, so we will take $\text{Re}(C)\simeq\text{Im}(C)=a$. Then the constraint equations are
\begin{subequations}\label{bdc}
\begin{align}
 |\text{Re}(\Delta_d)-1|=a|(\text{Re}(g_{bd}^2)-\text{Im}(g_{bd}^2))| \\
 |\text{Im}(\Delta_d)|=a|(\text{Re}(g_{bd}^2)+\text{Im}(g_{bd}^2))|
\end{align}
\end{subequations}
Again from \cite{bbdata}, we take the following bounds (at $3\sigma$):
\begin{align}
 0.62\leq \text{Re}(\Delta_d)\leq 1.36,  \hspace{0.5in} -0.39\leq\text{Im}(\Delta_d)\leq -0.02
\end{align}
Note that, for the $\text{Im}(\Delta_d)$, the SM value of 0 is slightly outside the $3\sigma$ allowed region, but we choose to disregard this discrepancy in setting the limits, since it is very small. Now, notice that $|\text{Re}(\Delta_d)-1|\leq0.38$, and $|\text{Im}(\Delta_d)|\leq 0.39$.  Putting these all together with Eq.\ref{bdc}, we get similar constraints for $\text{Re}(g_{bd}^2)$ and $\text{Im}(g_{bd}^2)$ such that $g_{bd}^2$ lies in an approximate circle of radius ${\sim}0.39/a$. So we get
\begin{subequations}
\begin{align}
& |\text{Re}(g_{bd}^2)|< 2.10\times 10^{-7}, \hspace{0.3in}  |\text{Im}(g_{bd}^2)|< 2.10\times 10^{-7} \\
&\hspace{0.6in}\Longrightarrow |g_{bd}| < 4.58\times 10^{-4}
\end{align}
\end{subequations}

\subsection{\texorpdfstring{$K^0-\bar{K}^0$} {K-meson} Mixing}

\begin{figure}[h]
\begin{center}
 \includegraphics[scale=0.8]{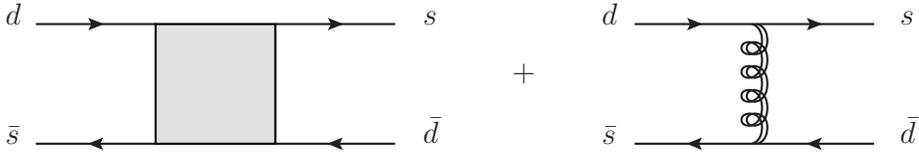}
 \caption{The SM and the axigluon contribution to $K-\bar{K}$ mixing}\label{kkmixing}
\end{center}
\end{figure}

The last constraint on $\epsilon^d$ comes from $K^0-\bar{K}^0$ mixing. The SM and the axigluon contributions to $K^0-\bar{K}^0$ mixing are shown in Fig.\ref{kkmixing}. The LO SM contribution again comes from EW box diagrams, but in this case $c$- and $t$-quark loops both need to be considered.  The NLO mixing amplitude from the SM is \cite{burasnlo}:
\begin{align}\label{smkbar1}
 \mathcal{M}_{SM}=-i\frac{G_F^2M_W^2}{16\pi^2}&\left( \lambda_c^2(1.86)S_0(x_c)+\lambda_t^2(0.574)S_0(x_t)+2\lambda_c\lambda_t(0.47)S_0(x_c,x_t)\right)
(\alpha_s(\mu))^{-2/9}\left[1+\frac{\alpha_s(\mu)}{4\pi}J_3\right](\bar{s}d)_{V-A}(\bar{s}d)_{V-A}
\end{align}
where 
\begin{align}
 S_0(x_c,x_t)&=x_c\left(\ln\frac{x_t}{x_c}-\frac{3x_t}{4(1-x_t)}-\frac{3x_t^2}{4(1-x_t)^2}\ln x_t\right)
\end{align}
is another Inami-Lin function, $\lambda_i=V_{is}^*V_{id}$, $J_3\simeq 1.895$, and $\mu\simeq \mathcal{O}(1$ GeV). The relevant matrix elements for kaon mixing can be found in \cite{kkdata}:
\begin{subequations} \label{mek}
\begin{align}
 &\langle\bar{K}^0|\bar{s}\gamma^\mu(1\pm\gamma_5)d\bar{s}\gamma^\mu(1\pm\gamma_5)d|K^0\rangle = {4\over3}m_{_{K^0}}f_K^2\hat{B}_K \\
 &\langle\bar{K}^0|\bar{s}\gamma^\mu(1+\gamma_5)d\bar{s}\gamma^\mu(1-\gamma_5)d|K^0\rangle = (7.8){8\over3}m_{_{K^0}}f_K^2\hat{B}_K \\
 &\langle\bar{K}^0|\bar{s}_\alpha\gamma^\mu(1+\gamma_5)d_\beta\bar{s}_\beta\gamma^\mu(1-\gamma_5)d_\alpha|K^0\rangle = (30.4){8\over3}m_{_{K^0}}f_K^2\hat{B}_K
\end{align}
\end{subequations}
Using Eq.\ref{mek} in Eq.\ref{smkbar1} we get
\begin{align} \label{smkbar2}
 \langle \bar{K}^0|\mathcal{H}_{SM}|K^0\rangle=\frac{G_F^2M_W^2}{12\pi^2}m_{_{K^0}}f_K^2\hat{B}_K\left[\lambda_c^2(0.001)+\lambda_t^2(1.776)+2\lambda_c\lambda_t(0.001)\right]
\end{align}

The axigluon contribution to kaon mixing can be written using Eq.\ref{baxmix} with the substitution of  appropriate quark operators. Then using Eq.\ref{mek} the axigluon matrix element becomes:
\begin{align}
 \langle\bar{K}^0|\mathcal{H}_{ax}|K^0\rangle& = \frac{m_{_{K^0}} f_K^2\hat{B}_K}{M_A^2}\left\{-{2C_0\over9}((g_{ds}^L)^2+(g_{ds}^R)^2)+\left(-1.73\, C_1+10.13\,C_2\right)g_{ds}^Lg_{ds}^R\right\} \notag \\
 &\simeq \frac{m_{_{K^0}} f_K^2\hat{B}_K}{M_A^2}\left\{-0.16((g_{ds}^L)^2+(g_{ds}^R)^2)+48.68\,g_{ds}^Lg_{ds}^R\right\}
\end{align}

Assuming again axial couplings: $g_{ds}^L=-g_{ds}^R=g_{ds}$, the axigluon matrix element is:
\begin{align}
 \langle\bar{K}^0|\mathcal{H}_{ax}|K\rangle = \frac{m_{_{K^0}} f_K^2\hat{B}_K}{4M_A^2}(48.36)g_{ds}^2
\end{align}

Now, following a common notation \cite{pdg}, we define
\begin{align}\label{m12}
 M_{12}=\langle\bar{K}^0|\mathcal{H}_{eff}|K^0\rangle
\end{align}
This matrix element $M_{12}$ is the off-diagonal element of the ``mass matrix" in the full Hamiltonian $H=M+{i\over2}\Gamma$, and $\mathcal{H}_{eff}$ is the effective Hamiltonian, that includes both the SM and the NP interactions, for the mixing process. The off-diagonal elements of $M$ are related to the mass difference between heavy and light mesons, and the off-diagonal elements of $\Gamma$ are related to the decay of these mesons. The interested reader should refer to \cite{pdg} and references therein for more information on meson mixings and decays. The CP violation in meson mixings comes from a possible phase difference between $M_{12}$ and $\Gamma_{12}$, which depends only on the short distance part of the matrix element $M_{12}$. The long distance interactions, which come from on-shell states in the loops (Fig.\ref{ewbox}), are CP conserving. The axigluon does not 
contribute to the long distance part of this amplitude at LO. The mass difference, $\Delta m$, between heavy and light mesons is  
\begin{align}
 \Delta m= 2|M_{12}|
\end{align}
Notice that $\Delta m$ gets affected by both the short distance and the long distance parts of the effective Hamiltonian. Unfortunately, calculation of the long distance contributions is difficult \cite{burasnlo}. In this paper we assume that long distance contributions are at most $50\%$ of the total mass difference, hence $M_{12}^{SM}=2M_{12}^{SD}$.

In order to constrain the coupling constant $g_{ds}$, we use the mass difference \cite{pdg} and the imaginary part of $M_{12}$ \cite{kkdata}:
\begin{align}
 \text{Im}(M_{12}^{NP})&=(1.7\pm1.6)\times 10^{-18}\, \text{GeV} \\
 \Delta m &= (3.483\pm 0.006)\times10^{-15}\, \text{GeV}
\end{align}

Consequently, we get the following constraints:
\begin{subequations}
\begin{align}
& |\text{Re}(g_{ds}^2)|<9.64\times 10^{-10}, \hspace{0.3in}  |\text{Im}(g_{ds}^2)|< 6.13\times 10^{-13} \\
&\hspace{0.6in}\Longrightarrow |g_{ds}| < 3.11\times 10^{-5}
\end{align}
\end{subequations}

Finally, we can write $\epsilon^d$ constraints as follows:
\begin{align}\label{fccdown}
|\text{Im}(\epsilon^d)|<\left(\begin{array}{ccc}  
         0 & 9.88{\times}10^{-9} & 2.08{\times}10^{-4}\\
	 9.88{\times}10^{-9} & 0 & 7.69{\times}10^{-4} \\
	 2.08{\times}10^{-4} & 7.69{\times}10^{-4} & 0
         \end{array} \right), \hspace{0.5in} 
|\epsilon^d|< \left(\begin{array}{ccc}  
         0 & 0.311 & 4.58\\
	 0.311 & 0 & 18.3 \\
	 4.58 & 18.3 & 0
         \end{array} \right)\times 10^{-4}
\end{align}
where $\text{Im}(\epsilon^d)$ is calculated assuming $\text{Im}(g_{ij})<\text{Re}(g_{ij})$ and $\text{Im}(g_{ij}),\text{Re}(g_{ij})\neq0$.


\subsection{\texorpdfstring{$D^0-\bar{D}^0$}{D-meson} Mixing}

\begin{figure}[b]
\begin{center}
 \includegraphics[scale=0.8]{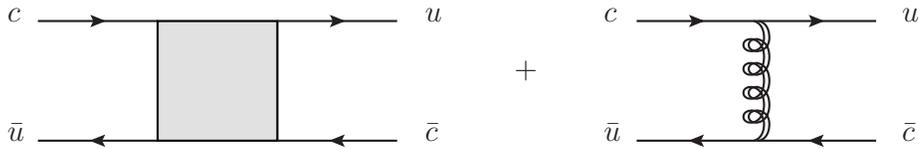}
 \caption{The SM and the axigluon contribution to $D^0-\bar{D}^0$ mixing}\label{ddmixing}
\end{center}
\end{figure}

$D^0-\bar{D}^0$ mixing, like the other meson mixings, happens via EW box diagrams in the SM, shown in Fig.\ref{ddmixing}. Calculation of the LO SM contribution to the mixing amplitude is similar to $K^0-\bar{K}^0$ mixing where $c-$ and $t-$loops are changed with  $s$- and $b$-loops. $D^0-\bar{D}^0$ mixing suffers more from long distance contributions as compared to $K^0$ and $B^0$ mixing. Consequently, NLO corrections to the mixing amplitude are not calculated in the literature. In this section, we only consider the LO short distance amplitude, and assume that long distance effects are at most the same order as short distance ones. The LO short distance SM contribution to the mixing amplitude (at $\mu=2$ GeV) is: 
\begin{align}
 \mathcal{M}_{SM}&=-i\frac{G_F^2 M_W^2}{16\pi^2}(0.80)\left\{ \lambda_s^2S_0(x_s)+\lambda_b^2S_0(x_b) +2\lambda_s\lambda_b S_0(x_s,x_b)  \right\} (\bar{u}c)_{V-A}(\bar{u}c)_{V-A} \\
&\simeq -i\frac{G_F^2 M_W^2}{16\pi^2}\left\{ \lambda_s^2\,(1.24\times10^{-6})+\lambda_b^2\,(2.18\times10^{-3}) +2\lambda_s\lambda_b\,(9.23\times10^{-6})  \right\} (\bar{u}c)_{V-A}(\bar{u}c)_{V-A}
\end{align}
For the matrix elements, we assume similar relations to Eq.\ref{meb} since $m_{_{D^0}}\simeq m_c + m_u\simeq 1.86$ GeV. The axigluon contribution can be written using Eq.\ref{baxmix} with the substitution of the appropriate quark operators: 
\begin{align}
 \langle \bar{D}^0|\mathcal{H}_{SM}|D^0\rangle\simeq=&\frac{G_F^2 M_W^2}{12\pi^2}\left\{ \lambda_s^2\,(1.24\times10^{-6})+\lambda_b^2\,(2.18\times10^{-3}) +2\lambda_s\lambda_b\,(9.23\times10^{-6})  \right\} m_{D^0}f_{D^0}^2 \\
\langle \bar{D}^0|\mathcal{H}_{ax}|D^0\rangle\simeq&\frac{1}{M_A^2}\left\{-0.18 ((g_{uc}^R)^2+(g_{uc}^L)^2)-1.04\,g_{uc}^Rg_{uc}^L\right\}m_{D^0}f_{D^0}^2
\end{align}

Once again, we assume axial couplings $g_{uc}^R=-g_{uc}^L=g_{uc}$:
\begin{align}
 \langle \bar{D}^0|\mathcal{H}_{ax}|D^0\rangle\simeq\frac{m_{D^0}f_{D^0}^2}{M_A^2}(0.68)g_{uc}^2
\end{align}

Constraints on $g_{uc}$ come from the $D^0-\bar{D}^0$ mass difference, $\Delta m_D=1.57\times10^{-14}$ GeV \cite{pdg}, and the ratio $\frac{q}{p}$, where $q$ and $p$ are the coefficients that describe the flavor eigenstates $D^0, \bar{D}^0$ in terms of mass eigenstates $D^0_H, D^0_L$:
\begin{align*}
 |D_L\rangle=p|D^0\rangle +q|\bar{D}^0\rangle \\
 |D_H\rangle=p|D^0\rangle -q|\bar{D}^0\rangle
\end{align*}
One can show that
\begin{align}
 {q\over p}=\sqrt{\frac{M_{12}^*-{i\over2}\Gamma_{12}^*}{M_{12}-{i\over2}\Gamma_{12}}}
\end{align}
In $D^0-\bar{D}^0$ system, $\Gamma_{12}\simeq M_{12}$ \cite{pdg}, and so we have
\begin{align}\label{qp}
 {q\over p}=\frac{2M_{12}^*}{\Delta m_D}
\end{align}
We use Eq.\ref{qp} to constrain the imaginary part of $g_{uc}$. The real part is already more constrained from the mass difference. From \cite{pdg}, we take the following values for the magnitude and the argument of $\frac{q}{p}$:
\begin{align}
 \left|{q\over p}\right|=0.60 \hspace{0.5in}  \text{Arg}\left({q\over p}\right)=-22.1^\circ
\end{align}
The constraints on $g_{uc}$ from these values are as follows:
\begin{subequations}
\begin{align}
& |\text{Re}(g_{uc}^2)|<2.09\times 10^{-8}, \hspace{0.3in}  |\text{Im}(g_{uc}^2)|< 4.89\times 10^{-9} \\
&\hspace{0.6in}\Longrightarrow |g_{uc}| < 1.47\times 10^{-4}
\end{align}
\end{subequations}

Unfortunately, there are no other mesons with which we can investigate the up-sector further. However, the neutral $D$-meson system has other interesting features, like the CP asymmetry in $D^0\to h^+h^-$ decays that was measured in 2011 \cite{lhcb-D}. We look more into the contribution of FC axigluons to this process in the next section.

\section{Some Examples of Flavor Changing Axigluon Contributions to Meson Decays}

In this section, we will check the effects of the FC couplings on several SM processes, namely  the rare decay $B^0_s\to\mu^+\mu^-$, the CP asymmetry in $D^0\to h^+h^-$ decays, and the isospin violation in $B^{0(+)}\to K^{(*)}\mu^+\mu^-$ decays. We do not expect a significant contribution to $B^0_s\to\mu^+\mu^-$ decay from the axigluons, since it is affected through axigluon-penguin loops at LO. However, when FC axigluon currents exist, processes like $D^0\to h^+h^-$ and $B^{0(+)}\to K^{(*)}\mu^+\mu^-$ can happen at tree level. Therefore one would expect to get an appreciable contribution from axigluon induced channels. These decays are chosen because they are of current experimental interest  \cite{LHCb-Bmar,LHCb-Bnov,lhcb-D,iso3}. 

\subsection{Rare decay \texorpdfstring{$B_s\to\mu^+\mu^-$}{Bs-> mu+mu-}}

In the SM, this decay is predicted to happen very rarely, with a branching ratio of $(3.5\pm0.30)\times10^{-9}$ \cite{Fleischer}. This is very close the recently measured branching ratio of $3.2^{+1.5}_{-1.2}\times10^{-9}$ by LHCb \cite{LHCb-Bmar,LHCb-Bnov}. These new results constrain the NP one can have that would affect this branching ratio. As we will see in this section, axigluon contribution to this decay amplitude is at least 3 orders of magnitude smaller than the SM contribution. Consequently, the measured branching ratio data does not impose further constraints on the coupling constant $g_{bs}$. 

This branching ratio is so small in the SM as it occurs through EW penguin and box diagrams. The SM effective Hamiltonian, including NLO corrections, is calculated in \cite{burasnlo} as:
\begin{align} \label{bmusm1}
 \mathcal{H}_{SM}(B_s^0\to\mu^+\mu^-)=i\frac{G_F}{\sqrt{2}}\frac{\alpha}{2\pi\sin^2\theta_W}(V_{tb}^*V_{ts}) Y(x_t)(\bar{s}b)_{V-A}(\bar{\mu}\mu)_{V-A} 
\end{align}
where
\begin{align}
Y(x_t)&=Y_0(x_t)+\frac{\alpha_s}{4\pi}Y_1(x_t) \\
Y_0(x)&= \frac{x}{8}\left(\frac{4-x}{1-x}+\frac{3x}{(1-x)^2}\ln x\right)\\
Y_1(x)&=\frac{4x+16x^2+4x^3}{3(1-x)^2}-\frac{4x-10x^2-x^3-x^4}{(1-x)^3}\ln x+\frac{2x-14x^2+x^3-x^4}{2(1-x)^3}\ln^2x\notag\\
&\hspace{20pt}+\frac{2x+x^3}{(1-x)^2}L_2(1-x)+8x\left.\frac{\partial Y_0(x)}{\partial x}\right |_{x_\mu}\ln x_{\mu} \label{y1}
\end{align}
and $\theta_W$ is the Weinberg angle. In Eq.\ref{y1}, $x_{\mu}=\frac{\mu^2}{M_W^2}$ where $\mu$ is the renormalization scale ($\mu\sim\mathcal{O}( m_{_{B_s}})$). For $m_t=172.1$ GeV and $M_W=80.4$ GeV, $Y(x_t)\simeq \mathcal{O}(1)$, and Eq.\ref{bmusm1} can be written as: 
\begin{align} \label{bmusm2}
 \mathcal{H}_{SM}(B_s^0\to\mu^+\mu^-)\simeq i\frac{G_F}{\sqrt{2}}\frac{\alpha}{2\pi\sin^2\theta_W}(V_{tb}^*V_{ts})(\bar{s}b)_{V-A}(\bar{\mu}\mu)_{V-A} 
\end{align}

\begin{figure}[h]
\begin{center}
 \includegraphics[scale=0.8]{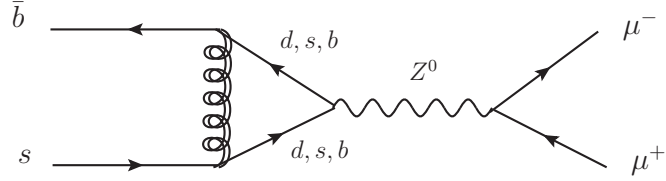}
 \caption{The axigluon contribution to $B^0_s\to\mu^+\mu^-$}\label{bmuaxf}
\end{center}
\end{figure}
The axigluon contribution to this decay is shown in Fig.\ref{bmuaxf}, which is a penguin loop with an axigluon. This is the LO contribution, since the axigluon does not couple to leptons. In this diagram, $b-$ and $s-$loops are more important than the $d-$loop, since the $d-$loop is doubly suppressed by the small coupling constant. One might worry about the divergence of the corresponding loop integral. This divergence arises because the part of the theory we are considering is not complete, for example there are additional (heavy) quarks in the full theory, and their inclusion should lead to cancellations of these divergences. The full theory is renormalizable \cite{tavares}, so we do not worry about the terms that can be canceled through the short distance contributions of the heavy quarks. Hence the amplitude (for the $b-$ quark loop) is: 
\begin{align}
  \mathcal{M}_{ax}= -\frac{G_F\tilde{g}_s}{4\pi^2}(\mu\bar{\mu})_{V-A}\,\bar{s}\left[\g^\nu(g_{bs}^RP_R+g_{bs}^LP_L)[(-v_b\g_5+a_b)B_1(y_b)+y_b(v_b\g_5+a_b)B_2(y_b)]\right]b
\end{align}
where
\begin{align}
B_1(y)&={1\over4}+\frac{1}{2(y-1)}+\frac{y(y-2)}{2(y-1)^2}\ln y\\
B_2(y)&=\frac{1}{y-1}-\frac{\ln y}{(y-1)^2}
\end{align}
and $y_q=\frac{m_q^2}{M_A^2}$. Adding the $s-$ loop, and realizing that $yB_2(y)<<B_1(y)$ for $y<<1$, we neglect the part ${\sim}B_2(y)$, and write as an $\mathcal{O}(1)$ estimate
\begin{align}
  \mathcal{M}_{ax}\simeq -\frac{G_F\tilde{g}_s}{4\pi^2}(B_1(y_b)+B_1(y_s))(\mu\bar{\mu})_{V-A}\,[\bar{s}\g^\nu(V-A\g_5)b]
\end{align}
where 
\begin{align*}
 V&=g_{bs}^Va_b-g_{bs}^Av_b \\
 A&=g_{bs}^Vv_b-g_{bs}^Aa_b
\end{align*}
together with $g^V_{bs}=\frac{g_{bs}^R+g_{bs}^L}{2}$ and $g^A_{bs}=\frac{g_{bs}^R-g_{bs}^L}{2}$. In this paper, we take $g_{bs}^V=0$. Thus the axigluon contribution to the $B^0_s\to\mu^+\mu^-$ decay hamiltonian becomes
\begin{align}\label{bmuax2}
 \mathcal{H}_{ax}= i\frac{G_F\tilde{g}_s\,g_{bs}}{4\pi^2}(B_1(y_b)+B_1(y_s))(\bar{s}b)_{V-A}\,(\mu\bar{\mu})_{V-A}
\end{align}

Now we can compare Eq.\ref{bmuax2} and Eq.\ref{bmusm2}, for $M_A=400$ GeV (at $M_W$ scale):
\begin{align}
 \frac{\langle\mathcal{H}_{ax}\rangle}{\langle\mathcal{H}_{SM}\rangle}\simeq0.001
\end{align}
The ratio gets smaller for smaller axigluon masses. Since the axigluon loop contributions are very small compared to the SM, the uncertainties in the calculations should not be a worry. Hence flavor changing axigluon couplings under already considered constraints do not affect the $B^0_s\to\mu^+\mu^-$ branching ratio in a noticeable way, and so this decay does not give further constraints on the coupling constant $g_{bs}$.

\subsection{CP Violation in \texorpdfstring{$D^0\to h^+h^-$}{D->h+h-} decays}
\begin{figure}[h]
\begin{center}
 \includegraphics[scale=0.8]{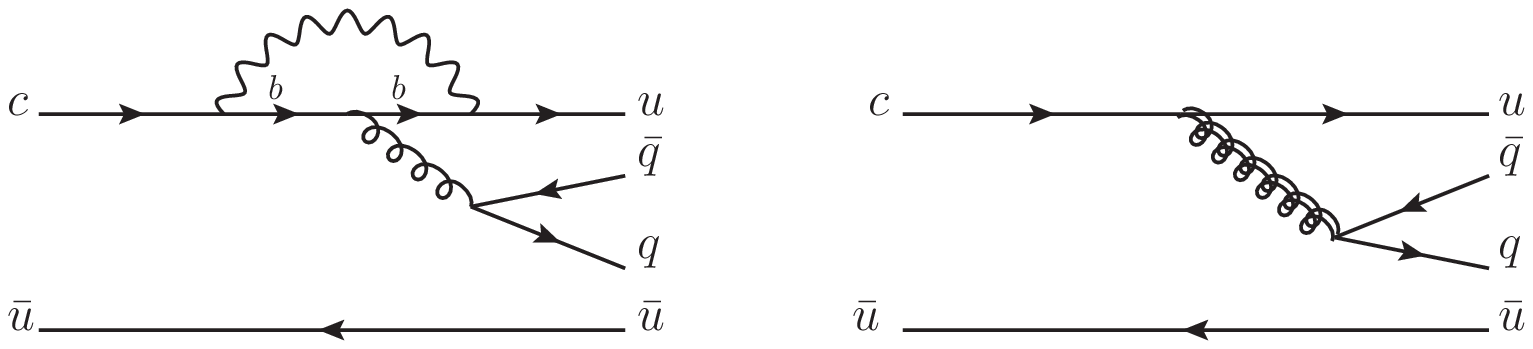}
 \caption{CP violation in $D^0\to h^+h^-$}\label{dcp}
\end{center}
\end{figure}
The CP asymmetry in $D^0\to h^+h^-$ decays is defined as follows:
\begin{align}
 A_{CP}=\frac{\Gamma(D^0\to h^+h^-)-\Gamma(\bar{D}^0\to h^+h^-)}{\Gamma(D^0\to h^+h^-)+\Gamma(\bar{D}^0\to h^+h^-)}
\end{align}
where $\Gamma(i\to f)$ is the partial width of the $i\to f$ decay. In 2011, the LHCb measured this asymmetry to be ${\sim}1\%$. In Moriond 2013, the LHCb presented a new, smaller measurement \cite{CPnew}. There still appears to be much confusion about the origin of this asymmetry in the SM \cite{Golden,Cheng,Grossman}. The long distance effects in $D$-decays make the calculation of relevant hadronic matrix elements very difficult. Furthermore, the charm quark might not be heavy enough to trust perturbation theory at this scale.  Naively, one would expect the CP violation at LO to come from decay channels like the first diagram in Fig.\ref{dcp}, which is estimated in \cite{Cheng,Grossman} to give ${\sim}0.1\%$. The discrepancy between $1\%$ (experiment) and $0.1\%$ (SM) might be due to new physics. However it also might be contained in the SM if some matrix elements of penguin operators are much larger than the estimates given by dimensional analysis \cite{Golden}. In this section, we compare the 
contributions of the two diagrams in Fig.\ref{dcp} to the CP asymmetry. 

The effective Hamiltonian that comes from the SM diagram in Fig.\ref{dcp} can be written as follows \cite{Buras}:
\begin{align}\label{cpsm}
 \mathcal{H}_{SM}=\frac{G_F}{\sqrt{2}}\lambda_b\left\{ C_{3/5}(\bar{u}c)_{V-A}(\bar{d}\gamma^\mu d)+C_{4/6}(\bar{u}_\alpha c_\beta)_{V-A}(\bar{d}_\beta\gamma^\mu d_\alpha) \right\}
\end{align}
where
\begin{align}
 C_{3/5}(M_W)=-{1\over3}C_{4/6}(M_W)=-\frac{\alpha_s(M_W)}{24\pi}\bar{E}_0(x_b)
\end{align}
and 
\begin{align}
 \bar{E}_0(x)=-\frac{2}{3}\ln(x)+\frac{x(18-11x-x^2)}{12(1-x)^3}+\frac{x^2(15-16x+4x^2)}{6(1-x)^4}\ln(x)-{2\over3}
\end{align}
is another Inami-Lin function. The subscripts of the coefficients and the choice of writing the quark operators in this way is a slight variation of what Buras does in his paper \cite{Buras}. He gathers operators and their coefficients in a way that is easier to keep track of in the RG flow equations. Here we do not RG flow the coefficients, and compare the SM and the axigluon parts at $M_W\simeq M_A\simeq\mathcal{O}(100$ GeV). The effective axigluon Hamiltonian that can be written from the diagram on the right in Fig.\ref{dcp} is:
\begin{align} \label{cpax}
 \mathcal{H}_{ax}=\frac{g_{uc}\tilde{g}_s}{M_A^2}\left\{-{1\over6}(\bar{u}\gamma^\mu\gamma_5c)(\bar{d}\gamma_\mu\gamma_5d)+{1\over2}(\bar{u}_\alpha\gamma^\mu\gamma_5c_\beta)(\bar{d}_\beta\gamma_\mu\gamma_5d_\alpha)\right\}
\end{align}
where $\tilde{g}_s\simeq \frac{g_s}{3}$ \cite{tavares}. Comparing Eq.\ref{cpax} with Eq.\ref{cpsm} for $M_A=400$ GeV, we see that the upper bound for axigluon contribution is an order of magnitude larger than the SM contribution. This upper bound grows with decreasing axigluon mass, to ${\sim}40$ times the SM contribution for $M_A=100$ GeV. Thus this could produce larger than the SM CP violation in $D$-meson decays.

\subsection{Isospin violation in \texorpdfstring{$B\to K^{(*)}\mu^+\mu^-$} {B->Kmu+mu-} decays}

\begin{figure}[h]
 \begin{center}
  \includegraphics[scale=0.8]{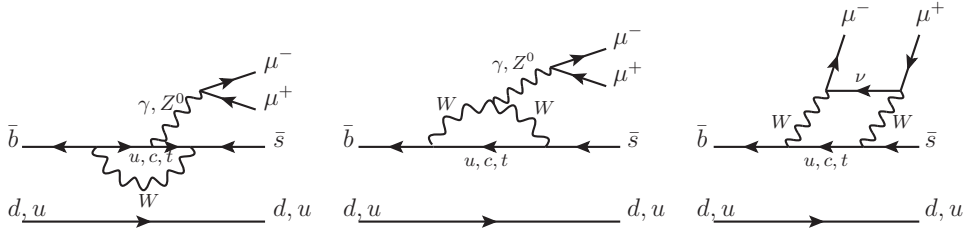}
  \caption{Isospin conserving diagrams in the SM for $B^{0(+)}\to K^{(*)}\mu^+\mu^-$ decays} \label{nonisosm}
 \end{center}
\end{figure}

\begin{figure} [h]
\centering
\subfloat[One of the isospin violating SM diagrams]{
\includegraphics[scale=0.8]{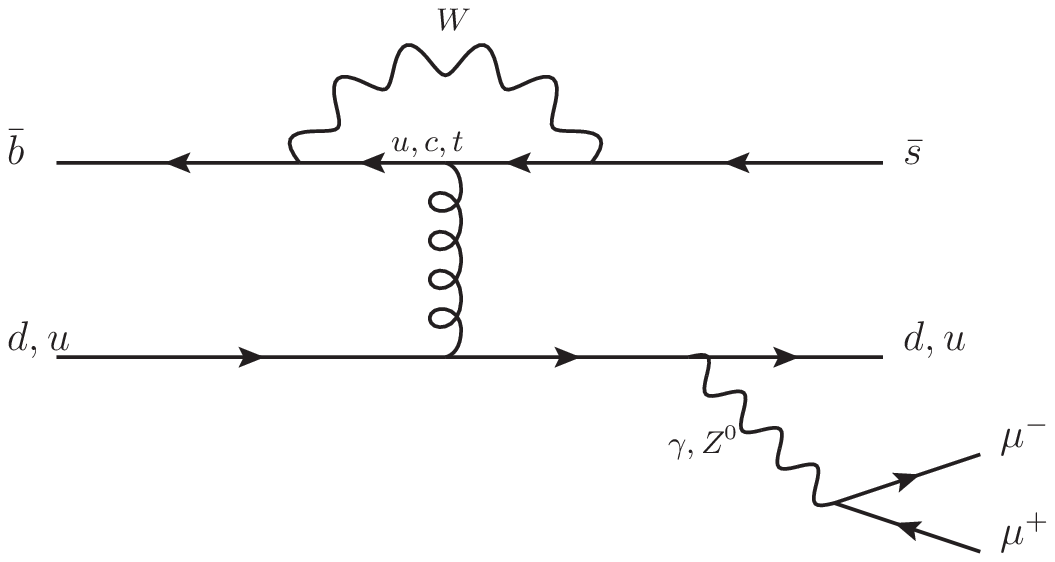}
\label{isosm}
}
\subfloat[Axigluon contribution to isospin violation]{
\includegraphics[scale=0.8]{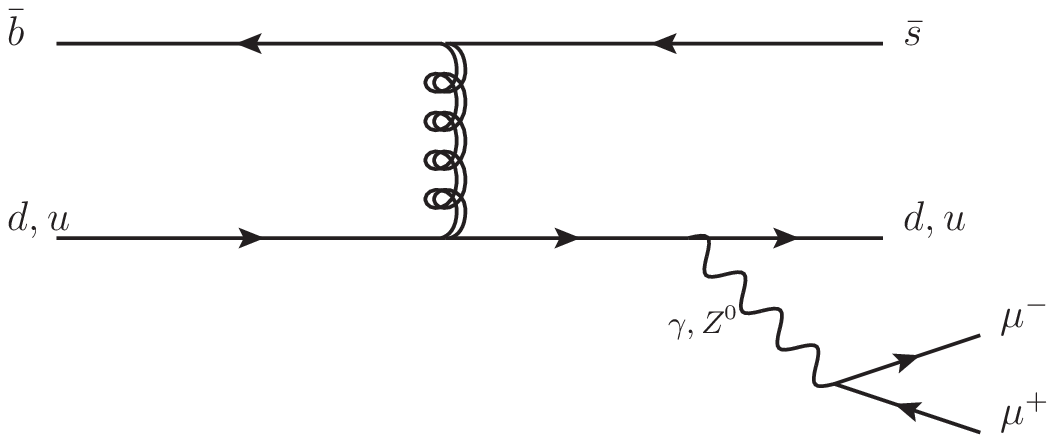}
 \label{isoax}
}
\caption{Isospin violation in $B^{0(+)}\to K^{0(+)}\mu^+\mu^-$ decay} \label{iso}
\end{figure}

In the SM, $B\to K \mu^+\mu^-$ decay follows, at LO, from EW penguin and box diagrams that do not involve the $d(u)$-quark, which is then called the spectator quark (Fig.\ref{nonisosm}).  For this decay, an observable, the isospin asymmetry $A_I$, can be defined as follows:
\begin{align}
 A_I=\frac{\Gamma(B^0\to K^0 \mu^+\mu^-)-\Gamma(B^+\to K^+ \mu^+\mu^-)}{\Gamma(B^0\to K^0 \mu^+\mu^-)+\Gamma(B^+\to K^+ \mu^+\mu^-)}
\end{align}
A similar asymmetry is defined also for $B\to K^*\mu^+\mu^-$. We can see that in the spectator quark approximation, this asymmetry is zero, since there is no difference between the decay of the neutral and charged $B-$meson. If we consider diagrams in which the final $\mu^+\mu^-$ pair is emitted from the spectator quark (Fig.\ref{iso}), there would be a non-zero isospin asymmetry due to the different charges of the spectator quarks involved in neutral and charged $B-$meson decays. In the SM, the asymmetry for $B\to K^*\mu^+\mu^-$ is expected to be around $-1\%$ \cite{iso1,iso2}. Although there is no clear prediction for the isospin asymmetry in $B\to K\mu^+\mu^-$  from the SM, one might expect it to be similarly small, almost zero \cite{iso3}. The isospin asymmetry that is measured at the LHCb is consistent with the SM for the $B\to K^*\mu^+\mu^-$, however it deviates from zero with $4.4\sigma$ significance for $B\to K\mu^+\mu^-$ \cite{iso3}.
 The SM prediction might be enhanced by more precise hadronic matrix element calculations, however there might as well be NP involved in these decays, like flavor changing axigluons (Fig.\ref{isoax}).

In order to compare the axigluon contribution to this isospin violating process with the SM contribution, we assume that the EW penguin diagram in Fig.\ref{isosm} is as important as any other isospin violating diagram in the SM, if not the most important one. Therefore, instead of performing comprehensive calculations of the SM contributions, we only compare the two diagrams that are shown in Fig.\ref{iso}. Furthermore, we only consider the parts of these diagrams that are responsible for $B^0\to K^0$ decay, since the emission of the final muons are the same in both cases.

The SM contribution to the hadronic part of the effective amplitude from Fig.\ref{isosm} can be written as follows:
\begin{align} \label{isosm1}
 \mathcal{M}_{SM}\simeq\lambda_i\frac{G_F}{\sqrt{2}}\frac{\alpha_s}{24\pi}E_0(x_i) (\bar{b}s)_{V-A}(\bar{d}d)
\end{align}
where $i=u,c,t$.

The axigluon part of the same amplitude from Fig.\ref{isoax} is:
\begin{align}\label{isoax1}
 \mathcal{M}_{ax}\simeq\frac{g_{bs}\tilde{g_s}}{6M_A^2}(\bar{b}s)_{V-A}(\bar{d}d)
\end{align}

Comparing Eq.\ref{isosm1} and Eq.\ref{isoax1} for $M_A=400$ GeV (at $M_W$ scale), we get

\begin{align}
 \left|\frac{\langle\mathcal{H}_{ax}\rangle}{\langle\mathcal{H}_{SM}\rangle}\right|\simeq0.3
\end{align}
The axigluon contribution is at most the same order as the SM one when $M_A=100$ GeV.

\section{Conclusion}
The light axigluon model is an experimentally allowed modification of the SM and a viable explanation of the CDF $t\bar{t}$ forward-backward asymmetry. In this paper we used the axigluon model suggested in \cite{tavares} to predict $b\bar{b}$ and $c\bar{c}$ forward-backward asymmetries. They are expected to be large and depend on the invariant mass of the quark pair. This mass dependence is a useful tool to investigate the mass of the axigluon.

We also modified this flavor conserving axigluon model to include flavor violating couplings between the axigluon and the SM quarks. These couplings are constrained by neutral meson mixings, and the upper bounds on their magnitudes are in the range $10^{-3}-10^{-5}$. After taking the upper bounds for the couplings, we checked their effects on the rare decay $B^0_s\to\mu^+\mu^-$, the CP violation in  $D^0\to h^+h^-$, and the isospin violation in $B\to K^{(*)}\mu^+\mu^-$. We found that the FC axigluon has virtually no effect on the decay $B^0_s\to\mu^+\mu^-$, since this process still occurs via loop diagrams. This result agrees with the last measurements of the branching ratio of this decay \cite{LHCb-Bmar,LHCb-Bnov}. In the case of the isospin violation in $B\to K^{(*)}\mu^+\mu^-$, FC axigluon effects seem to be at most the same order as the SM ones even though the axigluon contribution is at tree level. The most interesting effect of the FC axigluon is on CP violation in $D^0\to h^+h^-$ decays. For this CP 
violating asymmetry, the upper bound on the axigluon contribution is at least 10 times larger than the ${\sim}0.1\%$ SM prediction. We conclude that adding small flavor violating effects to the light axigluon model might contribute to the CP violation in $D^0\to h^+h^-$ and to neutral meson mixings.

\begin{acknowledgments}
I would like to thank Ann Nelson for inspiring this work and for our many fruitful conversations. I would also like to thank Dan Amidei for conversations about $t\bar{t}$ and $b\bar{b}$ forward-backward asymmetry measurements at the Tevatron. This work was supported
in part by the U.S. Department of Energy under Grant No. DE-FG02-96ER40956.

After this work was completed \cite{Gresham} appeared, which also discusses light axigluon constraints.
\end{acknowledgments}


\end{document}